%% file: main.tex
\documentclass[lettersize,journal]{IEEEtran}
\usepackage{amsmath,amsfonts}
\usepackage{algorithmic}
\usepackage{algorithm}
\usepackage{array}
\usepackage[caption=false,font=normalsize,labelfont=sf,textfont=sf]{subfig}
\usepackage{textcomp}
\usepackage{stfloats}
\usepackage{url}
\usepackage{verbatim}
\usepackage{graphicx}
\usepackage{cite}
\usepackage{colortbl}
\usepackage{xcolor}

\begin{document}

\title{A $10.8mW$ Mixed-Signal Simulated Bifurcation Ising Solver using SRAM Compute-In-Memory with $0.6\mu s$ Time-to-Solution}

\author{Alana Marie Dee,~\IEEEmembership{Student Member,~IEEE,}
        Sajjad Moazeni,~\IEEEmembership{Member,~IEEE}
\thanks{This paper was produced by the IEEE Publication Technology Group. They are in Piscataway, NJ.}
\thanks{Manuscript received April 19, 2021; revised August 16, 2021.}}

\markboth{Journal of \LaTeX\ Class Files,~Vol.~14, No.~8, August~2021}%
{Shell \MakeLowercase{\textit{et al.}}: A Sample Article Using IEEEtran.cls for IEEE Journals}

\IEEEpubid{0000--0000/00\$00.00~\copyright~2021 IEEE}

\maketitle

\input{abstract}

\begin{IEEEkeywords}
Ising machine, SRAM compute-in-memory, MAXCUT, simulated bifurcation, mixed-signal, optimization.
\end{IEEEkeywords}

\input{introduction}
\input{architecture}
\input{circuits}
\input{results}
\input{comparison_table}
\input{conclusion}
\vspace{-0.5cm}

\bibliographystyle{IEEEtran}
\bibliography{ising-refs-2025}

\vspace{-0.35cm}
\section{Biographies}

\vspace{-1.2cm}

\begin{IEEEbiographynophoto}{Alana Marie Dee}
(Graduate Student Member, IEEE) received a B.S. degree in electrical engineering from the University of Pittsburgh, Pittsburgh, PA, USA, in 2021, with a focus on signal processing and communications. They are currently a Graduate Student Researcher at the University of Washington, Seattle, WA, USA. Their research interests include mixed-signal integrated circuits and systems for emerging compute paradigms. They received the 2022 Cadence Diversity in Technology Scholarship and the 2024 Clean Energy Institute Graduate Fellowship.
\end{IEEEbiographynophoto}

\vspace{-1.2cm}

\begin{IEEEbiographynophoto}{Sajjad Moazeni}
(Member, IEEE) received a B.S. degree in electrical engineering from the Sharif University of Technology, Tehran, Iran, in 2013, and M.S. and Ph.D. degrees in electrical engineering and computer science from the University of California at Berkeley, Berkeley, CA, USA, in 2016 and 2018, respectively. From 2018 to 2020, he was a Post-Doctoral Research Scientist with the Bioelectronic Systems Laboratory, Columbia University, New York, NY, USA. He is currently an Assistant Professor with the Electrical and Computer Engineering Department, University of Washington, Seattle, WA, USA. His research interests include designing integrated systems using emerging technologies, integrated photonics, neuro and bio photonics, and analog/mixed-signal integrated circuits. He received the 2022 NSF CAREER Award and the 2023 Google Faculty Award.
\end{IEEEbiographynophoto}

\end{document}

%% file: abstract.tex
\begin{abstract}

Combinatorial optimization problems are fundamental for various fields ranging from finance to wireless networks. This work presents a simulated bifurcation (SB) Ising solver in CMOS for NP-hard optimization problems. Analog domain computing led to a superior implementation of this algorithm as inherent and injected noise is required in SB Ising solvers. The architecture novelties include the use of SRAM compute-in-memory (CIM) to accelerate bifurcation as well as the generation and injection of optimal decaying noise in the analog domain. We propose a novel 10-T SRAM cell capable of performing ternary multiplication. When measured with 60-node, $50\%$ density, random, binary MAXCUT graphs, this all-to-all connected Ising solver reliably achieves above $93\%$ of the ground state solution in $0.6\mu s$ with $10.8mW$ average power in TSMC 180nm CMOS. Our chip achieves an order of magnitude improvement in time-to-solution and power compared to previously proposed Ising solvers in CMOS and other platforms.

\end{abstract}

%% file: introduction.tex
\section{Introduction}

\begin{figure}[htbp]
\vspace{-0.15cm}
\centerline{\includegraphics[width=3.5in]{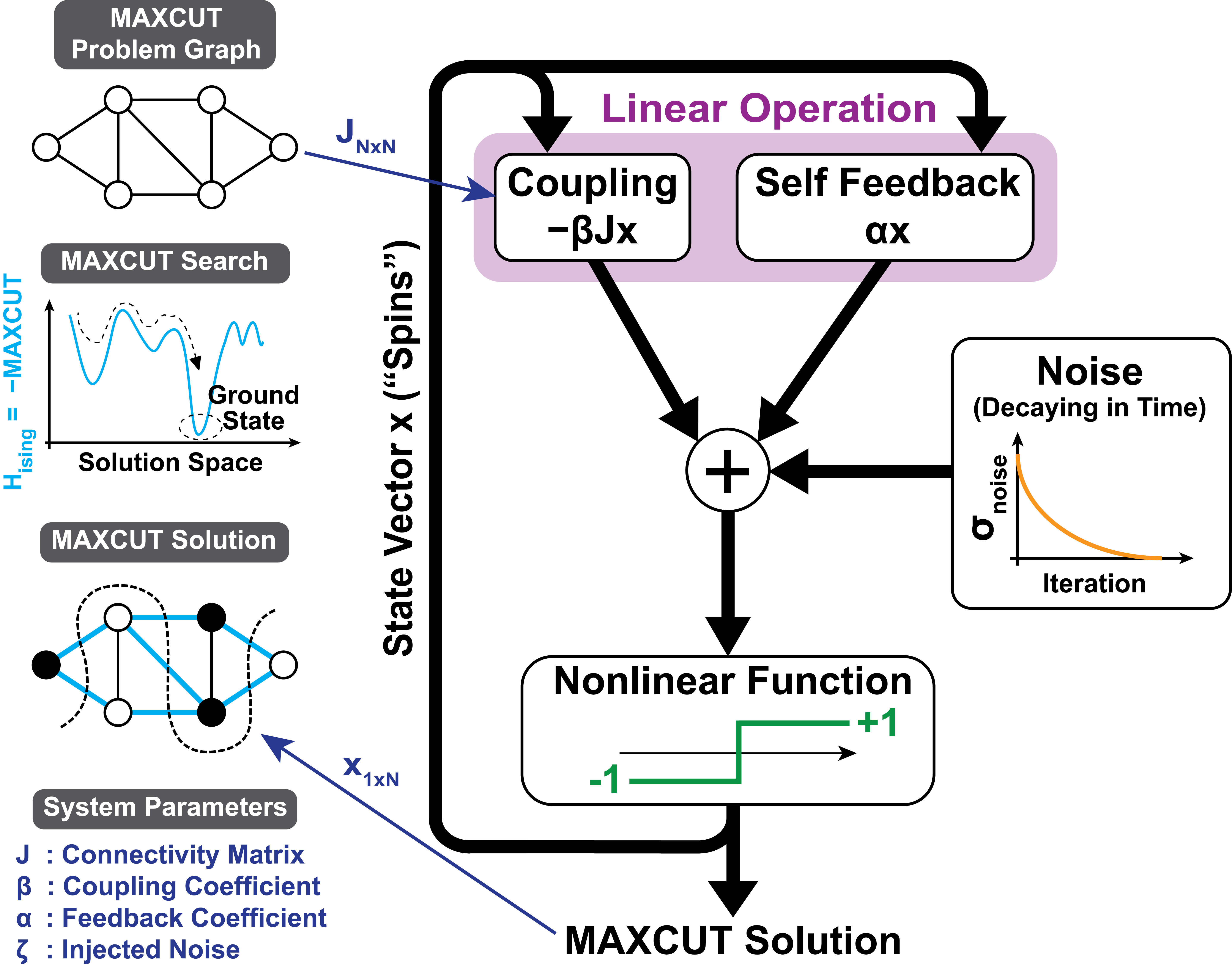}}
\vspace{-0.15cm}
\caption{Concept of a binary-weighted MAXCUT problem and the
proposed simulated bifurcation (SB) Ising machine.}
\label{fig:concept}
\vspace{-0.15cm}
\end{figure}

The Ising machine is an emerging quantum-inspired paradigm that accelerates solving NP-hard combinatorial optimization problems (COPs) such as MAXCUT. Practical applications require higher accuracy, lower run-times, and better scalability beyond approximate algorithms such as Goemans-Williamson (GW) or quantum computing~\cite{Bohm-2019-NatComm,Tatsumura-2021-NatElec,Kim-2023-ISSCC}. Real-time edge optimization applications such MIMO decoding~\cite{Kim-2021-MobiCom} require faster time-to-solution (TTS) as the problem size scales. Ising solvers map COPs to the Ising Hamiltonian and mimic quantum spin dynamics to perform a ground state search~\cite{Bohm-2019-NatComm,Xie-2022-JSSC}. Existing solvers utilize networks of coupled spins~\cite{Inagaki-2016-Science,Lo-2023-NatElec,Bae-2024-ISSCC}, stochastic neural networks (NNs)~\cite{Kim-2023-ISSCC}, or simulated annealing techniques~\cite{Tatsumura-2021-NatElec,Cai-2020-NatElec,Xie-2022-JSSC}. However, long convergence times ($>100\mu s$)~\cite{Tatsumura-2021-NatElec,Kim-2023-ISSCC,Inagaki-2016-Science} as well as poor energy efficiency~\cite{Tatsumura-2021-NatElec,Inagaki-2016-Science} make these approaches insufficient for optimization at the edge. Simulated bifurcation (SB) can reduce TTS and overcome scalability limitations of coupled oscillators~\cite{Lo-2023-NatElec} by emulating spins with a mathematically equivalent feedback loop shown in Fig.~\ref{fig:concept}~\cite{Bohm-2019-NatComm}. 

Additionally, techniques such as parallel spin updates~\cite{Bae-2024-ISSCC} and decaying noise~\cite{Cai-2020-NatElec} also accelerate TTS. We propose a mixed-signal SB Ising solver in CMOS~\cite{Dee-2024-TCAS} to target fast solution times and overcome shortfalls in ~\cite{Xie-2022-JSSC,Bae-2024-ISSCC,Cai-2020-NatElec} though all-to-all graph connectivity and an order-of-magnitude power reduction, respectively.

%% file: architecture.tex
\section{System Architecture}

This work introduces SRAM compute-in-memory (CIM) and analog noise generation in the SB operation to improve TTS by leveraging circuit variability and noise for this inherently analog system~\cite{Dee-2024-TCAS}. The solver’s digital backend provides a synchronous SB loop, high compatibility with digital processors, and fast reprogramming, addressing the slow operation and reprogramming of analog domain approaches such as memristive crossbars~\cite{Cai-2020-NatElec}. Random binary 60-node, $50\%$ density, all-to-all connected MAXCUT graphs were used as benchmark problems. An $N$-node graph is represented by an edge connectivity matrix, $\mathbf{J}$, and has solution, $\vec{x}$, encoding each node state. Cut size includes every edge connecting nodes of different states and is calculated through the Ising Hamiltonian~\cite{Bohm-2019-NatComm}. The SB process in Fig.~\ref{fig:concept} iteratively searches the solution space to converge on an optimal $\vec{x}$. Each SB cycle implements Eq.~\ref{eq1}, where $\vec{x}_{k+1}$ is influenced by self-feedback, coupling with $\mathbf{J}$, and iteration dependent generated noise, $\zeta_k$~\cite{Bohm-2019-NatComm}.

\begin{equation}
\vec{x}_{k+1} = \text{sgn}\left( \alpha \vec{x}_k - \beta\mathbf{J}\vec{x}_k + \zeta_k \right)
\label{eq1}
\end{equation}

\IEEEpubidadjcol

\begin{figure}[htbp]
\vspace{-0.15cm}
\centerline{\includegraphics[width=3.25in]{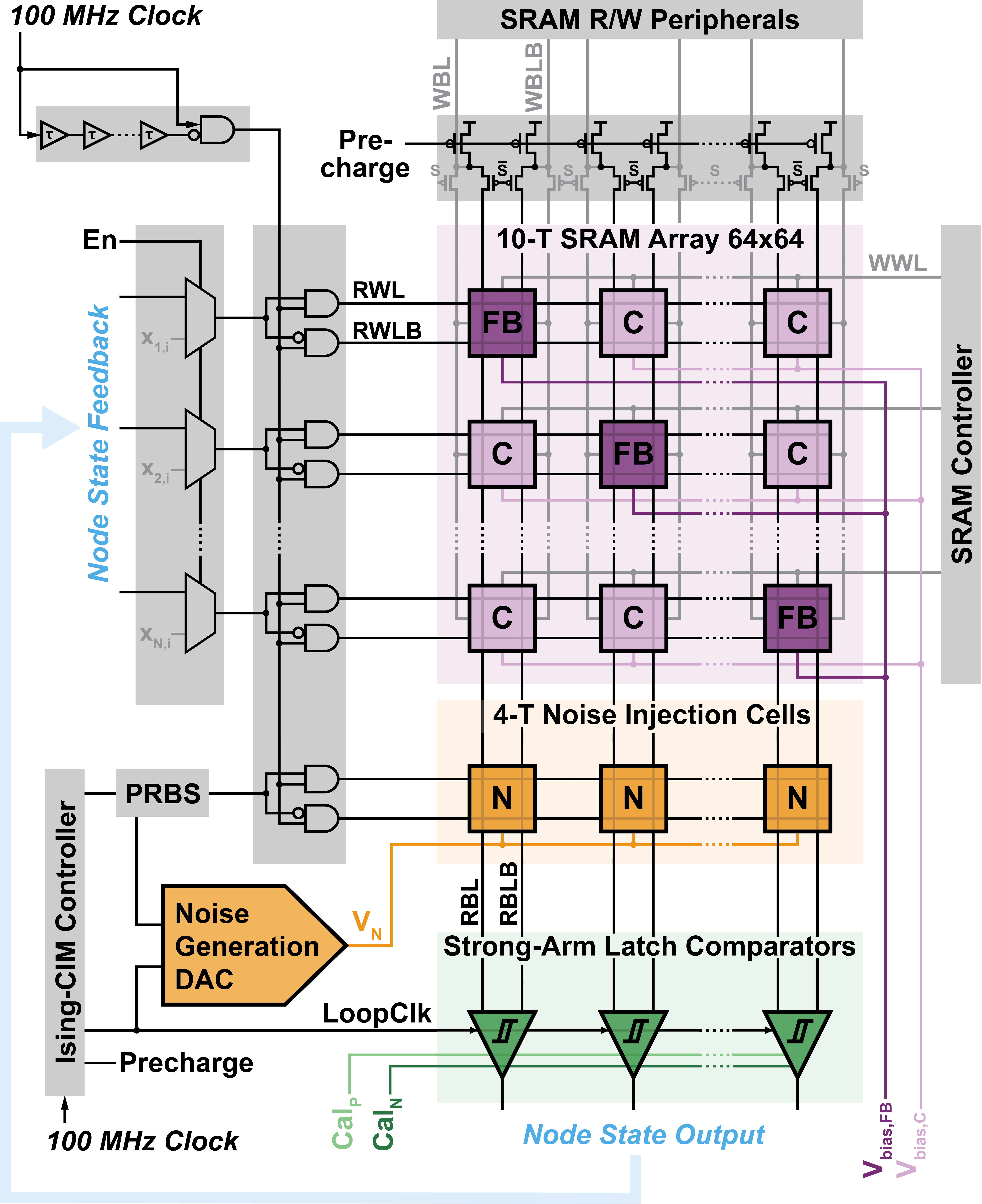}}
\vspace{-0.15cm}
\caption{Proposed mixed-signal architecture with custom cells for SB Ising-CIM and analog noise injection.}
\label{fig:system}
\vspace{-0.15cm}
\end{figure}

Decaying the power of $\zeta_k$ noise over successive cycles accelerates bifurcation~\cite{Cai-2020-NatElec}. Effective tuning control of loop parameters ($\alpha$, $\beta$, $\zeta$) in hardware is critical for optimal bifurcation. The addition of analog noise in this work allows for unique COPs to be solved with a single tuning point. Our SB circuit architecture in Fig.~\ref{fig:system} completes a loop iteration in $30ns$, controlled by a $100MHz$ reference clock. 

Prior to solver runs, the SRAM array is programmed to store $\mathbf{J}$ in coupling (C) cells and diagonal ones in feedback (FB) cells. In each SB iteration, a current domain multiply-and-accumulate (MAC) operation executes the linear operations of Eq.~\ref{eq1}. In comparison to charge domain CIM, current domain implementations provide higher cell density and less circuit complexity, while being more sensitive to $I_{DS}$ nonlinearity and process variations of access transistors~\cite{Chen-2021-JSSC}. However, SB utilizes binary spin states, making it robust to these nonlinearities. Two shared external bias voltages, $V_{bias,FB}$ and $V_{bias,C}$, are distributed across the array to implement $\alpha$ and $\beta$ control over SRAM cell currents and can be tuned to compensate for process variations. Further, mixed-signal SB solvers require a certain level of injected noise in each iteration to achieve higher accuracies in a shorter time~\cite{Cai-2020-NatElec}. The nonlinearities, process/mismatch variations, and noise which limit the precision of current domain CIM in NN applications are in fact beneficial in SB Ising computing~\cite{Dee-2024-TCAS}. 

%% file: circuits.tex
\section{Circuit Design}

\begin{figure}[htbp]
\vspace{-0.15cm}
\centerline{\includegraphics[width=3.5in]{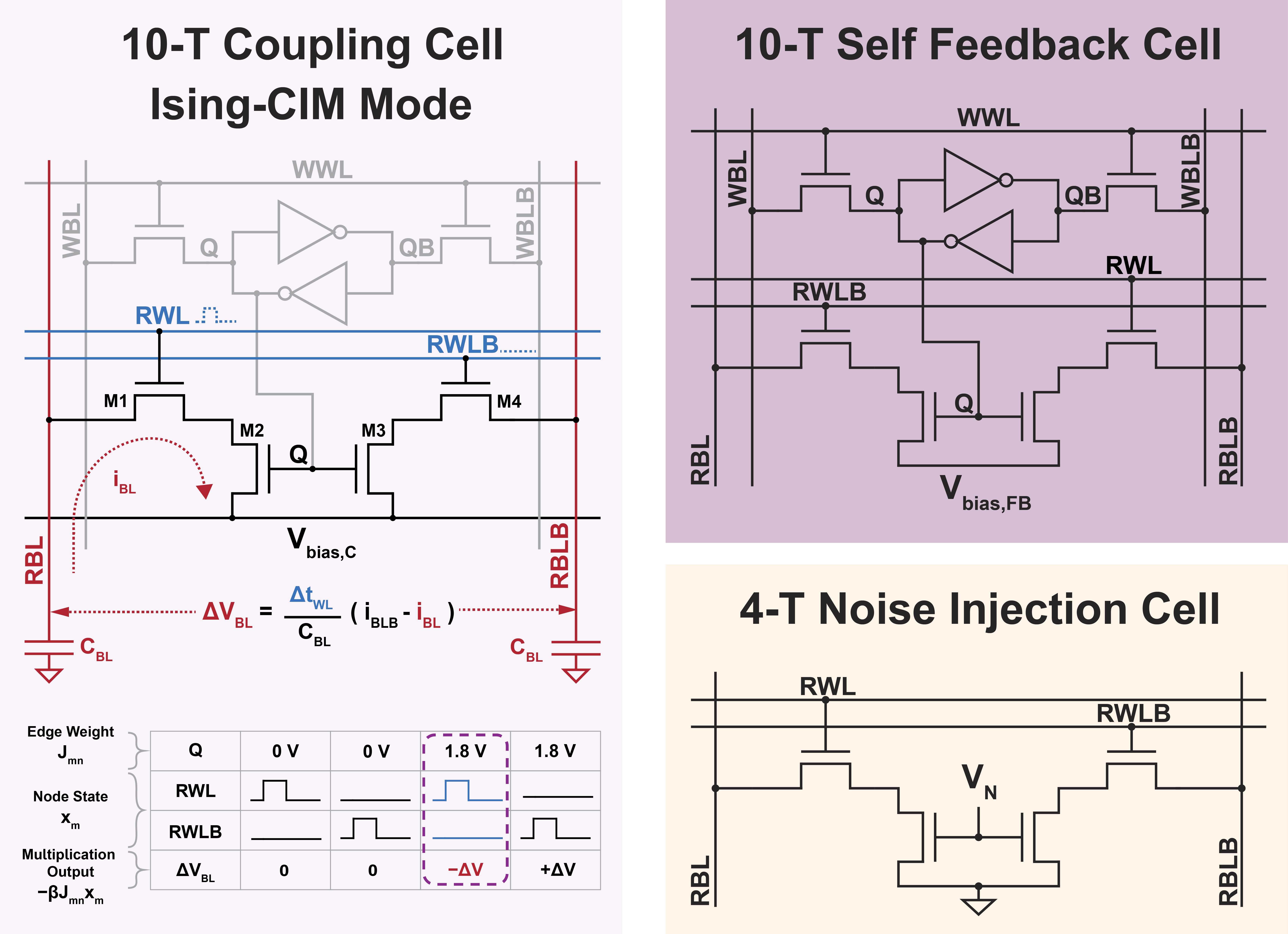}}
\vspace{-0.15cm}
\caption{Circuit implementation and SB operation mode of proposed coupling cell (left), custom self-feedback cell (top right), and custom current-domain noise injection cell (bottom right).}
\label{fig:cell}
\vspace{-0.15cm}
\end{figure}

The proposed 10-T SRAM macro in Fig.~\ref{fig:cell} implements a ternary multiplication specific to Eq.~\ref{eq1}, sinking bitline (BL) current when $J_{mn} = 1$ and $x_{m,k} = 1$, resulting in $x_{n,k+1} = -1$. 

\begin{figure}[htbp]
\vspace{-0.4cm}
\centerline{\includegraphics[width=2in]{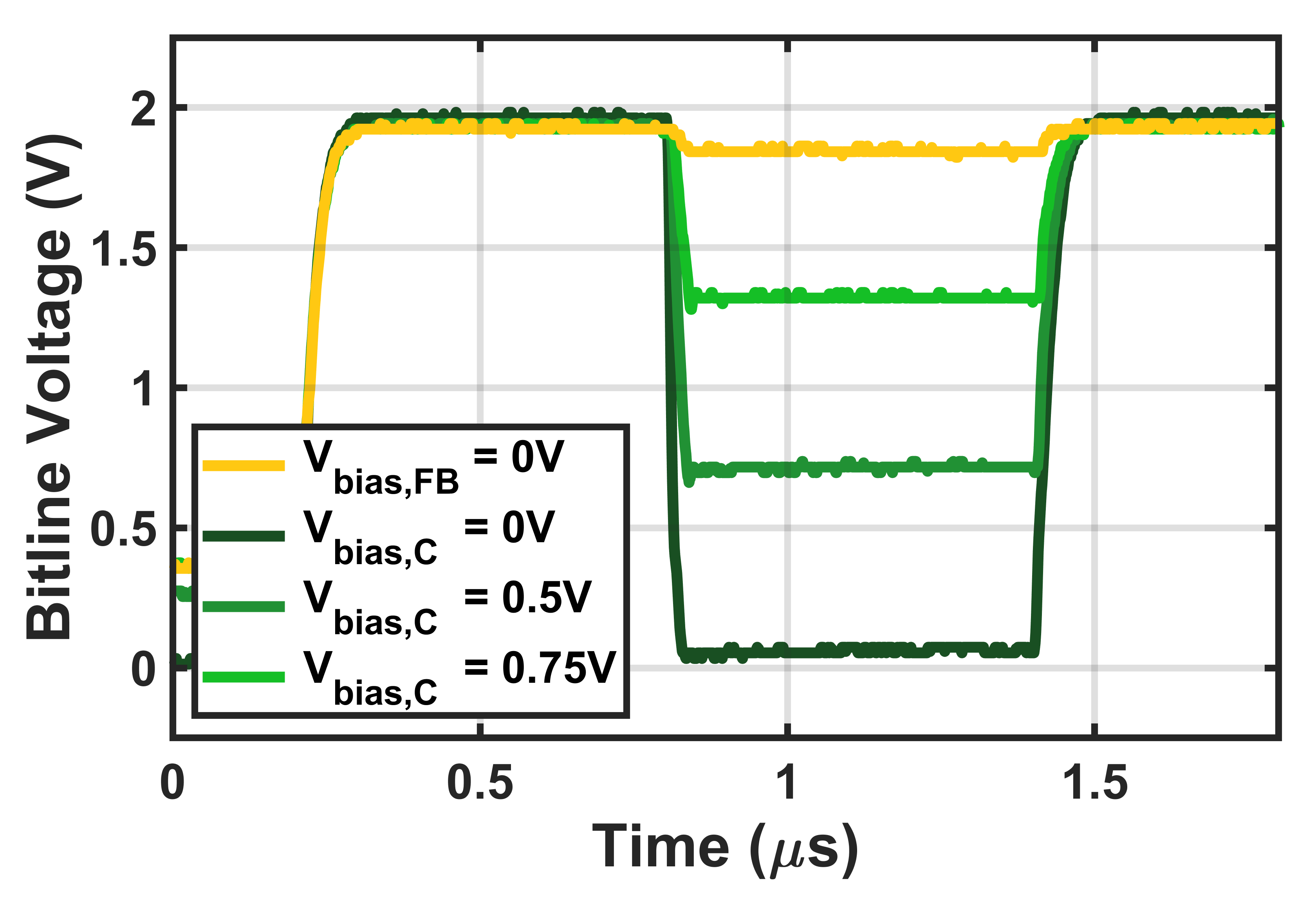}}
\vspace{-0.15cm}
\caption{Low-frequency circuit measurements of 10-T cell transient behavior.}
\label{fig:cellresults}
\vspace{-0.15cm}
\end{figure}

Differential wordlines (WLs) encode $x_{m,k}$ with $4ns$ pulses generated from a $100MHz$ clock, $1.8V$ VDD, and a tunable delay line with capacitively loaded buffers. Each C cell’s BL current contribution is proportional to the sizing of the access transistors M1/M4, $V_{bias,C}$ at the sources of M2/M3, and the WL pulse width and were designed to minimize effects of leakage current. FB cell operation is opposite, discharging BLB instead of BL when $x_{m,k} = 1$. This is scaled to the full array, where BL currents from a column’s C and FB cells are summed, executing a MAC operation. Cell behavior was experimentally verified in Fig.~\ref{fig:cellresults} by programming the array with ones. 

\begin{figure}[htbp]
\vspace{-0.15cm}
\centerline{\includegraphics[width=3.5in]{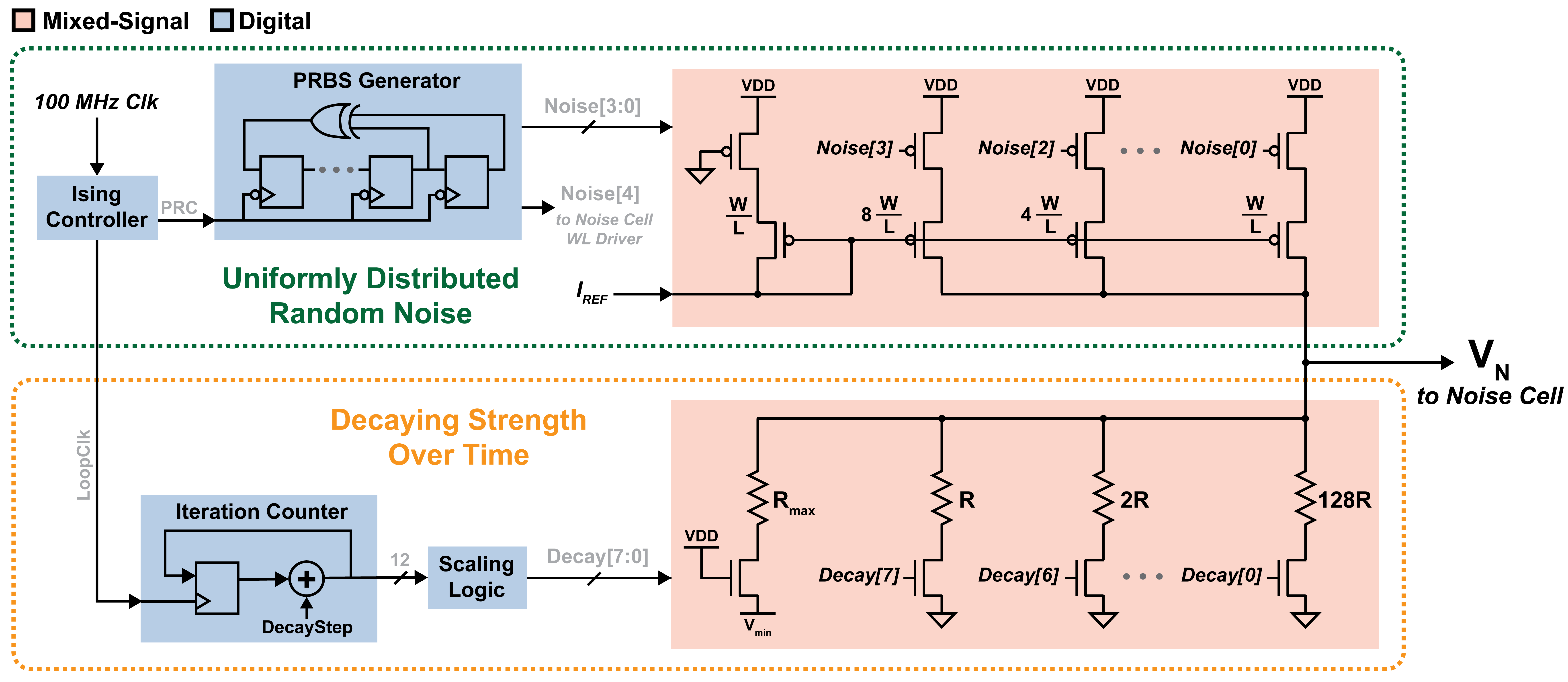}}
\vspace{-0.15cm}
\caption{Noise generation DAC for on-chip noise injection.}
\label{fig:noise}
\vspace{-0.15cm}
\end{figure}

The measured $V_{BL}$ and $V_{BLB}$ reflects a small voltage drop from the single FB cell and a large voltage drop from the 59 C cells. The drop in $V_{BLB}$ increases as $V_{bias,C}$ decreases, demonstrating control over the coupling strength, $\beta$. 
During each cycle, randomly varying current is added to the CIM output, perturbing the total BL current and implementing $\zeta$ in Eq.~\ref{eq1}. This is injected using the 4-T noise cell in Fig.~\ref{fig:cell}, where the gate voltage $V_N$ is generated with the on-chip DAC in Fig.~\ref{fig:noise} and the WL pulse polarity is determined by a psuedo-random bit sequence (PRBS). 

\begin{figure}[htbp]
\vspace{-0.5cm}
\centering
\subfloat[]{\includegraphics[width=1.5in]{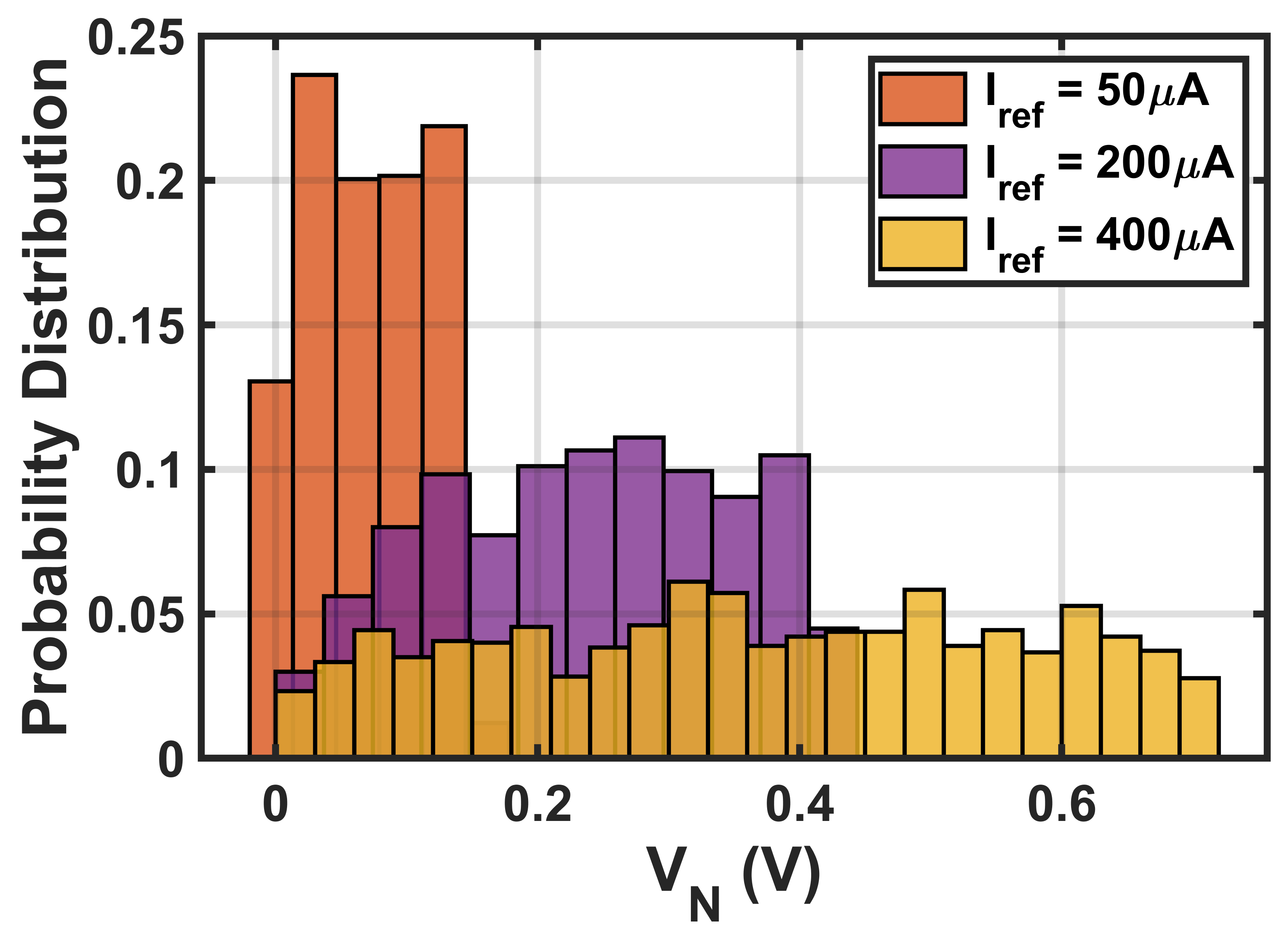}%
\label{fig_first_case}}
\hfil
\subfloat[]{\includegraphics[width=1.6in]{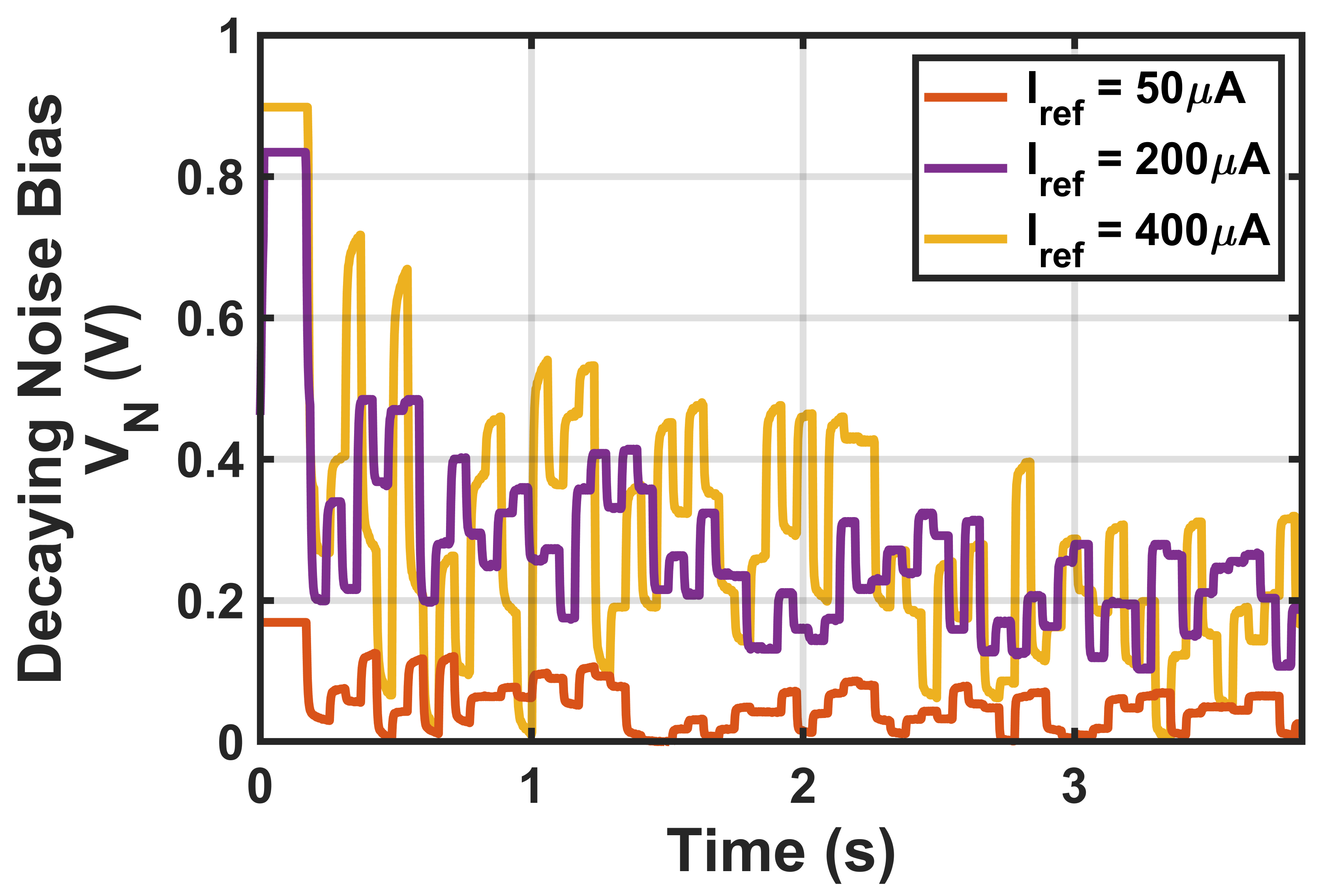}%
\label{fig_second_case}}
\caption{Low-frequency verification of on-chip noise generation DAC at different tuning current values. (a) Probability distribution without decay. (b) Decaying expected value.}
\label{fig:noiseresults}
\end{figure}

The expected current magnitude, controlled by an external bias, $I_{REF}$, has a uniform distribution that decays in magnitude each iteration. A current mirror DAC architecture, triggered by the precharge signal, uses 4 PRBS bits to drive binary-weighted current mirror branches, providing a random current value. A resistive DAC counterpart implements nonlinearly decaying noise. As the 12-bit iteration count increases, the adjusted 8-bit decay value increases and enables additional resistive branches. This decreases the effective resistance and consequently reduces $V_N$. Implementing randomness through current mirror branches ensures a uniform distribution for robust performance over different graphs. Decay was implemented with doubled resistive branches as this DAC nonlinearity further accelerates TTS. This behavior was verified using low frequency measurements of $V_N$ presented in Fig.~\ref{fig:noiseresults}.

The resulting BL current is the summation of self-feedback, coupling, and noise in Eq.~\ref{eq1} and discharges $200fF$ of pre-charged BL capacitance. This yields a unique $\Delta V_{BL}$ on each BL pair, which is thresholded by a strong-arm latch comparator (with an SR-latch as a dynamic-to-static converter at the output) at the rising edge of a third clock cycle. Strong-arm latch comparators have an extra input differential voltage shared across the row for offset calibration. The digitized states are fed back to the WL driver circuitry using multiple delay-matched routings for the next Ising iteration, prioritizing low latency and avoiding mismatch over area density.

%% file: results.tex
\section{Silicon Results}

\begin{figure}[htbp]
\vspace{-0.15cm}
\centerline{\includegraphics[width=2.7in]{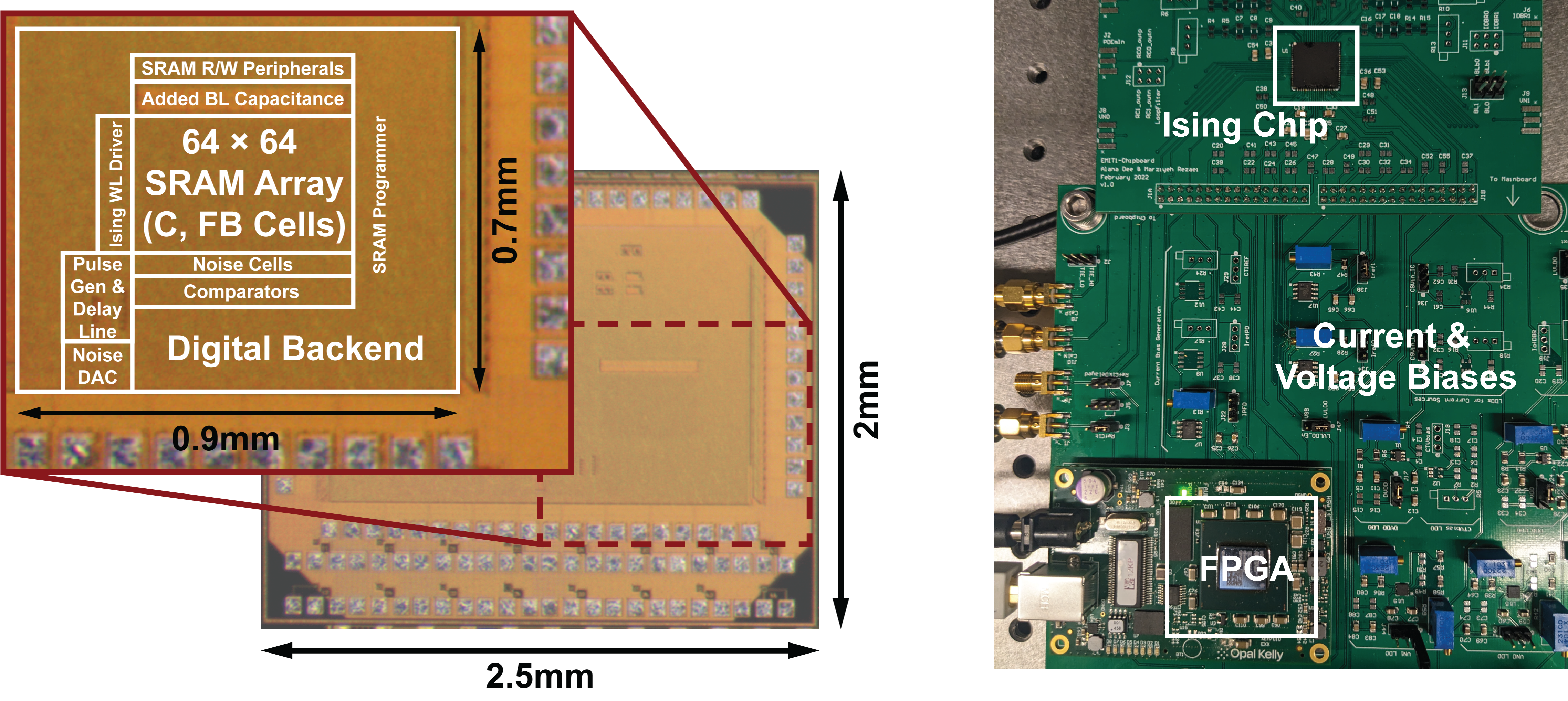}}
\vspace{-0.15cm}
\caption{Annotated chip micrograph (left) and experimental test setup with packaged chip and FPGA controller (right).}
\label{fig:chip}
\vspace{-0.15cm}
\end{figure}

In the presented room-temperature system measurements, self-feedback, coupling, and noise strength were controlled with $V_{bias,FB} = 0.57V$, $V_{bias,C} = 0.85V$, and $I_{REF} = 300\mu A$. Fig.~\ref{fig:chip} includes an annotated chip micrograph, indicating $\sim50\%$ area utilization alongside a picture of the test PCB infrastructure.

\begin{figure}[htbp]
\vspace{-0.3cm}
\centerline{\includegraphics[width=3in]{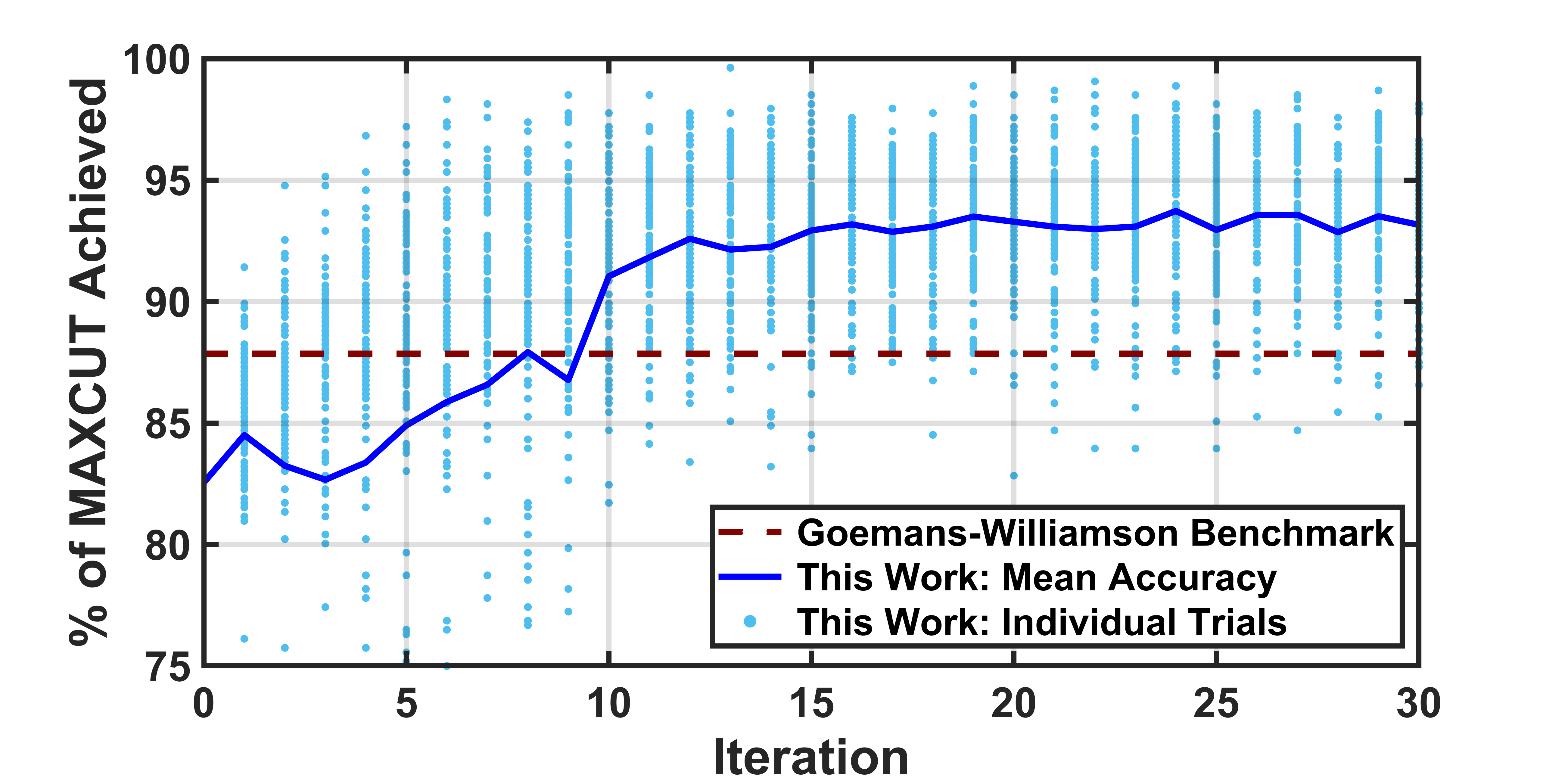}}
\vspace{-0.15cm}
\caption{Measured MAXCUT accuracy for a random graph (100 trials / iteration).}
\label{fig:results1}
\vspace{-0.15cm}
\end{figure}

An FPGA was used to interface with the Ising system’s serial scan chain interface and to provide a reference clock and voltage. The primary performance metric is MAXCUT accuracy, the ratio of the cut size found by this solver to the known ground state (GS) cut. For a single graph in Fig.~\ref{fig:results1}, the system converges to a constant accuracy within 20 iterations, or after $0.6\mu s$, and surpasses GW within $0.3\mu s$. 

\begin{figure}[htbp]
\vspace{-0.3cm}
\centerline{\includegraphics[width=2in]{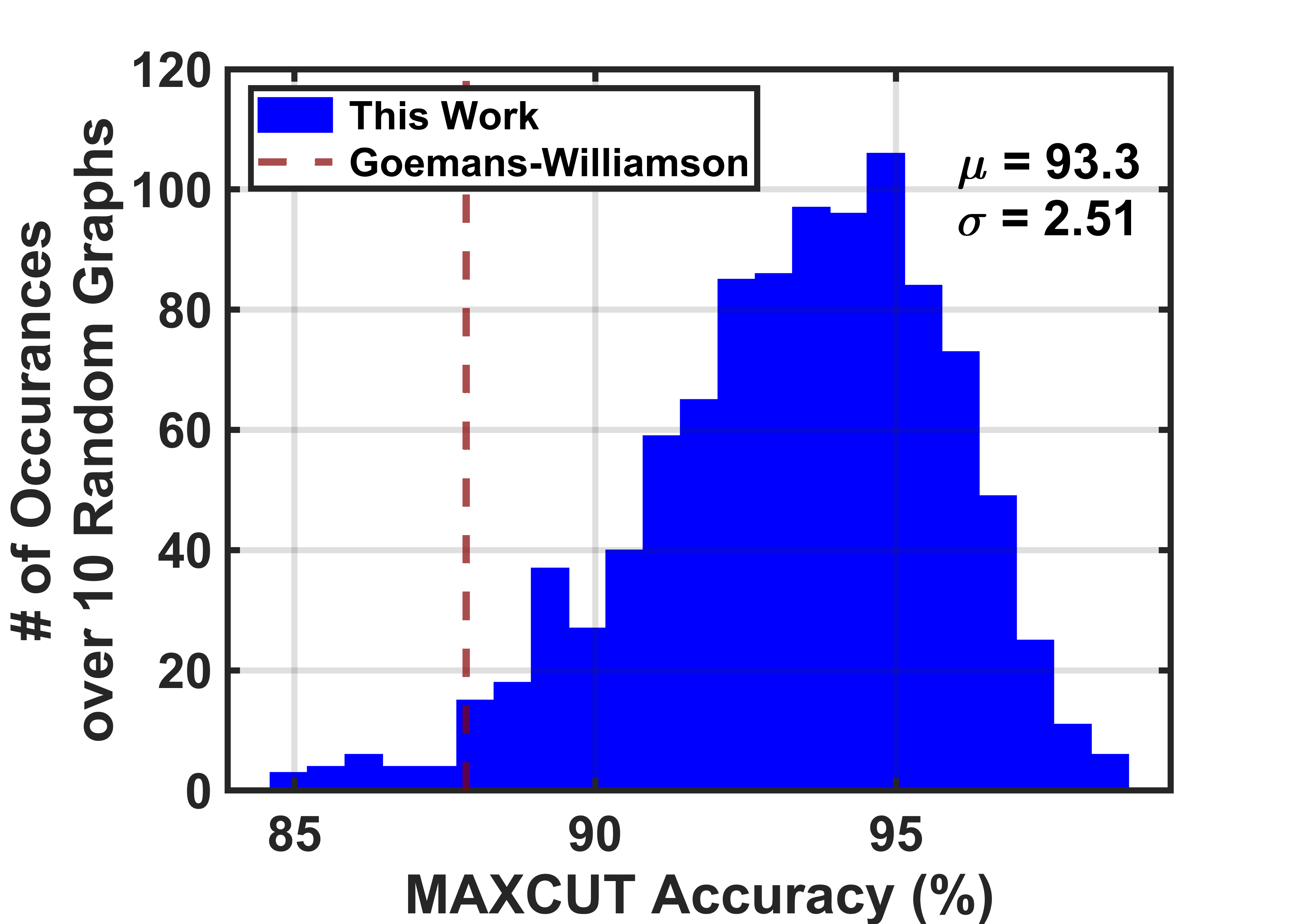}}
\vspace{-0.15cm}
\caption{Accuracy distribution after $0.6\mu s$ over a dataset of 10
random graphs (100 trials / graph / iteration).}
\label{fig:results2}
\vspace{-0.15cm}
\end{figure}

This solver was verified with 10 unique random graph topologies, where it is shown that a single bias point provides robust performance. The histogram in Fig.~\ref{fig:results2} shows that the average accuracy achieved after $0.6\mu s$ is $93.3\% ± 2.5\%$. 

\begin{figure}[htbp]
\vspace{-0.5cm}
\centerline{\includegraphics[width=2in]{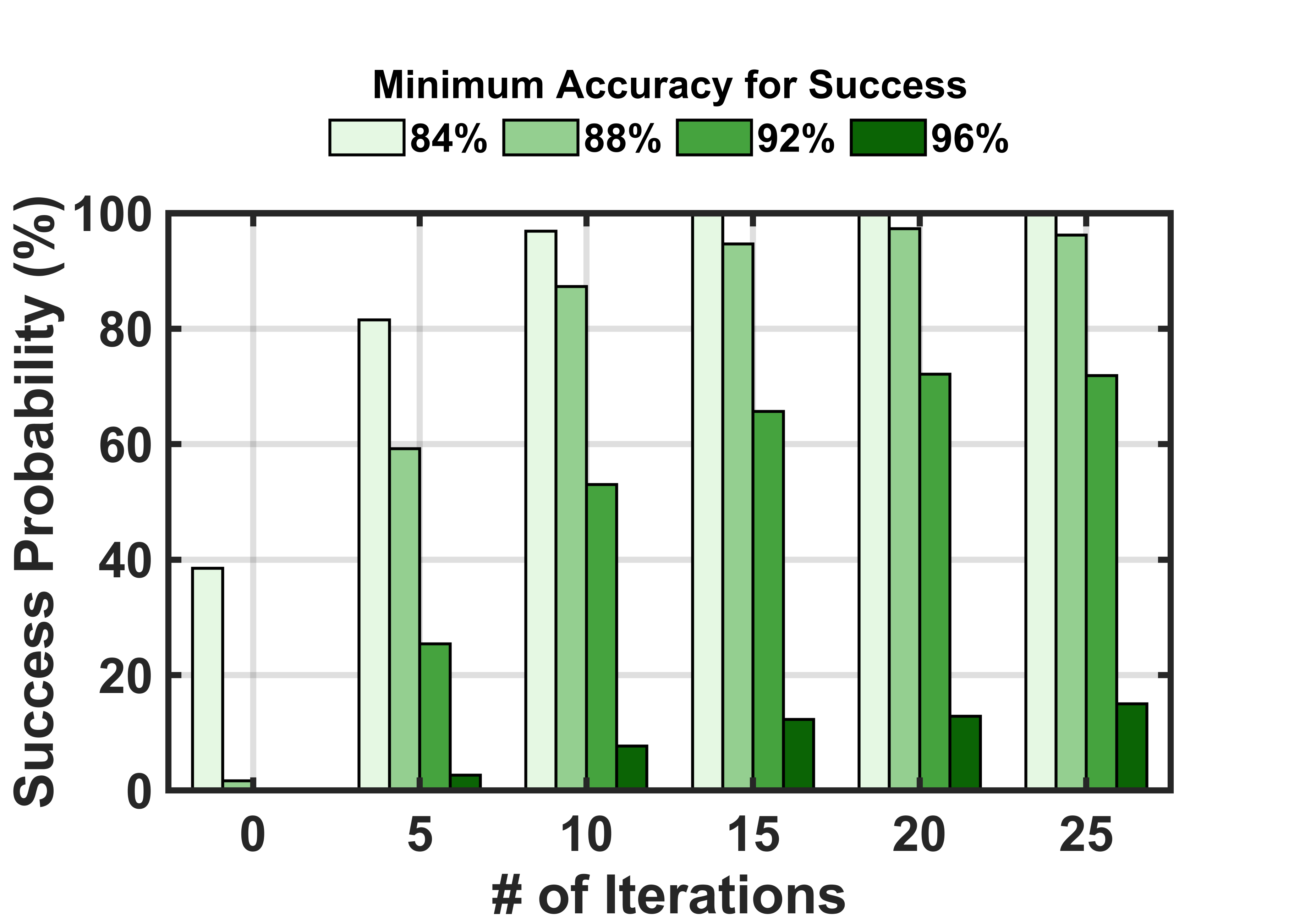}}
\vspace{-0.15cm}
\caption{Probability of achieving various accuracies over a dataset of 10
random graphs (100 trials / graph / iteration).}
\label{fig:results3}
\vspace{-0.15cm}
\end{figure}

This solver exceeds the accuracy guaranteed by GW in almost every trial. The success probability plot in Fig.~\ref{fig:results3} provides insights on the benefits of running the solver for more cycles. While 15 iterations provide good performance, an additional 5 iterations increase the probability of $+92\%$ accuracy from $66\%$ to $72\%$. 

This SB solver converges within $0.6\mu s$ for all-to-all connected NP-hard MAXCUT problems. 

%% file: comparison_table.tex
\begin{table*}[!t]
\caption{Comparison with State-of-the-Art Works.}
\resizebox{7.16in}{!}{%
\begin{tabular}{r c c c c c c}
\hline
\textbf{Work}                 & Nat. Elec. '23 & ISSCC '24 & JSSC '22 & Nat. Elec. '20 & Nat. Elec. '21 & \cellcolor{lightgray!40!white}\textbf{This}  \\ 

\textbf{}                 & \cite{Lo-2023-NatElec} & \cite{Bae-2024-ISSCC} &  \cite{Xie-2022-JSSC} &  \cite{Cai-2020-NatElec} & \cite{Tatsumura-2021-NatElec} & \cellcolor{lightgray!40!white} \textbf{Work} \\ 

\hline

\textbf{Solver Approach}                 & Coupled Ring & Coupled & Simulated & Simulated & Simulated & \cellcolor{lightgray!40!white} Simulated \\ 
\textbf{}                 & Oscillators & SRAM Spins & Annealing & Annealing & Bifurcation & \cellcolor{lightgray!40!white} Bifurcation  \\

\textbf{Compute-in-Memory Macro}         & N/A & N/A & eDRAM & Memristor & N/A & \cellcolor{lightgray!40!white} SRAM  \\ 

\textbf{Technology}     & 65nm                                          & 65nm                                     & 65nm                                & 16nm*                 & Multi-FPGA                      & \cellcolor{lightgray!40!white}180nm              \\ 

\textbf{Clock (MHz)}       & N/A                                           & N/A                                   & N/R                              & 500                            & 279                          & \cellcolor{lightgray!40!white}100             \\ 

\textbf{Number of Spins}      & 48                                            & 1536                                       & 6400                                 & 60**                           & 2000                              & \cellcolor{lightgray!40!white}60                 \\ 

\textbf{Input Graph Topology}      & all-to-all                                            & e-Chimera                                       & King's Graph                                & all-to-all                           & all-to-all                             & \cellcolor{lightgray!40!white} all-to-all                 \\ 
\textbf{}      & connected                                            & (11 neighbors)                                       & (8 neighbors)                                 & connected                           & connected                              & \cellcolor{lightgray!40!white} connected                 \\

\textbf{Convergence Time ($\mathbf{\mu} \text{s}$)} & 10                                          & 0.1                                    & 51.2                                 & 0.6                      & 220                           & \cellcolor{lightgray!40!white}0.6                \\ 

\textbf{Target Accuracy***}      & +95\%                                            & N/R                                       & N/R                                 & +99\%                           & N/R                              & \cellcolor{lightgray!40!white} +95\%       \\

\textbf{Area ($\text{mm}^2$)}       & 0.45                                   & 0.16\textdagger                                   & 0.71                               & 0.014                            & N/A                             & \cellcolor{lightgray!40!white}0.63               \\ 

\textbf{Power (mW)}       & 105                                           & 16.6                                     & 0.85                                 & 120.1                            & N/R                             & \cellcolor{lightgray!40!white}10.8\textdaggerdbl                \\ 
\hline
\end{tabular}
}
\\
\\N/A: Not Applicable, N/R: Not Reported
\\*Simulation results for parallel operation projected to the 16nm node. **Supports up to 1024 spins, but results correspond to 60 spins. ***For NP-hard random graph. \textdagger Core only, does not include digital backend. \textdaggerdbl $1.8V$ @ $100MHz$.
\label{comparison}
\vspace{-0.4cm}
\end{table*}

%% file: conclusion.tex
\section{Conclusion}

Table~\ref{comparison} shows that this work’s $10.8mW$ power consumption ($8.9mW$ dynamic power) and $0.63mm^2$ area is competitive, despite using an older technology node. We experimentally demonstrated sub-$\mu s$ TTS in 180nm. Other state-of-the art solutions with comprable TTS consume an order of magnitude more power per node~\cite{Cai-2020-NatElec} and have limited input graph connectivity~\cite{Bae-2024-ISSCC}. The novelty in this work of on-chip analog noise generation and injection alongside the application of SRAM CIM to SB allows for a solver that retains its performance across changes in required BL dynamic range and therefore can solve unique NP-hard input problems with a single tuning point. The system architecture can be implemented in more advanced processes to further improve the number of spins, area, speed, and energy consumption. The architecture of this novel mixed-signal Ising solver is suitable for high-speed, low-power edge applications such as massive MIMO detection inline acceleration for 5G/6G which require a solver with all-to-all edge connectivity. Future work will focus on demonstrating multi-bit edge weights and further scaling of problem size. 

%% file: main.bbl
\begin{thebibliography}{10}
\providecommand{\url}[1]{#1}
\csname url@samestyle\endcsname
\providecommand{\newblock}{\relax}
\providecommand{\bibinfo}[2]{#2}
\providecommand{\BIBentrySTDinterwordspacing}{\spaceskip=0pt\relax}
\providecommand{\BIBentryALTinterwordstretchfactor}{4}
\providecommand{\BIBentryALTinterwordspacing}{\spaceskip=\fontdimen2\font plus
\BIBentryALTinterwordstretchfactor\fontdimen3\font minus \fontdimen4\font\relax}
\providecommand{\BIBforeignlanguage}[2]{{%
\expandafter\ifx\csname l@#1\endcsname\relax
\typeout{** WARNING: IEEEtran.bst: No hyphenation pattern has been}%
\typeout{** loaded for the language `#1'. Using the pattern for}%
\typeout{** the default language instead.}%
\else
\language=\csname l@#1\endcsname
\fi
#2}}
\providecommand{\BIBdecl}{\relax}
\BIBdecl

\bibitem{Bohm-2019-NatComm}
\BIBentryALTinterwordspacing
F.~B{\"{o}}hm, G.~Verschaffelt, and G.~Van Der~Sande, ``{A poor man’s coherent Ising machine based on opto-electronic feedback systems for solving optimization problems},'' \emph{Nature Communications}, vol.~10, no.~1, 8 2019. [Online]. Available: \url{https://doi.org/10.1038/s41467-019-11484-3}
\BIBentrySTDinterwordspacing

\bibitem{Tatsumura-2021-NatElec}
\BIBentryALTinterwordspacing
K.~Tatsumura, M.~Yamasaki, and H.~Goto, ``{Scaling out Ising machines using a multi-chip architecture for simulated bifurcation},'' \emph{Nature Electronics}, vol.~4, no.~3, pp. 208--217, 3 2021. [Online]. Available: \url{https://doi.org/10.1038/s41928-021-00546-4}
\BIBentrySTDinterwordspacing

\bibitem{Kim-2023-ISSCC}
\BIBentryALTinterwordspacing
D.~Kim, N.~M. Rahman, and S.~Mukhopadhyay, ``{29.1 A 32.5mW Mixed-Signal Processing-in-Memory-Based k-SAT Solver in 65nm CMOS with 74.0\% Solvability for 30-Variable 126-Clause 3-SAT Problems},'' \emph{2022 IEEE International Solid- State Circuits Conference (ISSCC)}, pp. 28--30, 2 2023. [Online]. Available: \url{https://doi.org/10.1109/isscc42615.2023.10067570}
\BIBentrySTDinterwordspacing

\bibitem{Kim-2021-MobiCom}
\BIBentryALTinterwordspacing
M.~Kim, S.~Mandrà, D.~Venturelli, and K.~Jamieson, ``{Physics-inspired heuristics for soft MIMO detection in 5G new radio and beyond},'' \emph{Proceedings of the 28th Annual International Conference on Mobile Computing And Networking}, 3 2021. [Online]. Available: \url{https://doi.org/10.1145/3447993.3448619}
\BIBentrySTDinterwordspacing

\bibitem{Xie-2022-JSSC}
\BIBentryALTinterwordspacing
S.~Xie, S.~R.~S. Raman, C.~Ni, M.~Wang, M.~Yang, and J.~P. Kulkarni, ``{ISing-CIM: a reconfigurable and scalable compute within memory analog ISIng accelerator for solving combinatorial optimization problems},'' \emph{IEEE Journal of Solid-State Circuits}, vol.~57, no.~11, pp. 3453--3465, 6 2022. [Online]. Available: \url{https://doi.org/10.1109/jssc.2022.3176610}
\BIBentrySTDinterwordspacing

\bibitem{Inagaki-2016-Science}
\BIBentryALTinterwordspacing
T.~Inagaki, Y.~Haribara, K.~Igarashi, T.~Sonobe, S.~Tamate, T.~Honjo, A.~Marandi, P.~L. McMahon, T.~Umeki, K.~Enbutsu, O.~Tadanaga, H.~Takenouchi, K.~Aihara, K.-I. Kawarabayashi, K.~Inoue, S.~Utsunomiya, and H.~Takesue, ``{A coherent Ising machine for 2000-node optimization problems},'' \emph{Science}, vol. 354, no. 6312, pp. 603--606, 10 2016. [Online]. Available: \url{https://doi.org/10.1126/science.aah4243}
\BIBentrySTDinterwordspacing

\bibitem{Lo-2023-NatElec}
\BIBentryALTinterwordspacing
H.~Lo, W.~Moy, H.~Yu, S.~Sapatnekar, and C.~H. Kim, ``{An Ising solver chip based on coupled ring oscillators with a 48-node all-to-all connected array architecture},'' \emph{Nature Electronics}, vol.~6, no.~10, pp. 771--778, 8 2023. [Online]. Available: \url{https://doi.org/10.1038/s41928-023-01021-y}
\BIBentrySTDinterwordspacing

\bibitem{Bae-2024-ISSCC}
\BIBentryALTinterwordspacing
J.~Bae, C.~Shim, and B.~Kim, ``{15.6 e-Chimera: A Scalable SRAM-Based Ising Macro with Enhanced-Chimera Topology for Solving Combinatorial Optimization Problems Within Memory},'' \emph{2024 IEEE International Solid- State Circuits Conference (ISSCC)}, pp. 286--288, 2 2024. [Online]. Available: \url{https://doi.org/10.1109/isscc49657.2024.10454340}
\BIBentrySTDinterwordspacing

\bibitem{Cai-2020-NatElec}
\BIBentryALTinterwordspacing
F.~Cai, S.~Kumar, T.~Van~Vaerenbergh, X.~Sheng, R.~Liu, C.~Li, Z.~Liu, M.~Foltin, S.~Yu, Q.~Xia, J.~J. Yang, R.~Beausoleil, W.~D. Lu, and J.~P. Strachan, ``{Power-efficient combinatorial optimization using intrinsic noise in memristor Hopfield neural networks},'' \emph{Nature Electronics}, vol.~3, no.~7, pp. 409--418, 7 2020. [Online]. Available: \url{https://doi.org/10.1038/s41928-020-0436-6}
\BIBentrySTDinterwordspacing

\bibitem{Dee-2024-TCAS}
A.~M. Dee, D.~Vuong, K.~Bennett, and S.~Moazeni, ``Design of a mixed-signal compute-in-memory ising solver with sub-$\mu s$ time-to-solution and optimal decaying noise profile,'' \emph{IEEE Transactions on Circuits and Systems I: Regular Papers}, vol.~71, no.~12, pp. 5376--5386, 2024.

\bibitem{Chen-2021-JSSC}
\BIBentryALTinterwordspacing
Z.~Chen, Z.~Yu, Q.~Jin, Y.~He, J.~Wang, S.~Lin, D.~Li, Y.~Wang, and K.~Yang, ``{CAP-RAM: a Charge-Domain In-Memory computing 6T-SRAM for accurate and Precision-Programmable CNN inference},'' \emph{IEEE Journal of Solid-State Circuits}, vol.~56, no.~6, pp. 1924--1935, 5 2021. [Online]. Available: \url{https://doi.org/10.1109/jssc.2021.3056447}
\BIBentrySTDinterwordspacing

\end{thebibliography}
